# Development of Aluminium Based Surface Nano Composites Using Friction Stir Processing

Rahul R[1], K. V. Rajulapati[1], G. M. Reddy[2] and K. B. S. Rao[1]

[1]School of Engineering Sciences and Technology, University of Hyderabad, Hyderabad 500046 India.

[2]Defence Metallurgical Research Laboratory, Hyderabad 500005 India.

**Abstract**

Friction Stir Processing (FSP) is a relatively new technique which has been developed for microstructural modification of metallic materials through intense, localized plastic deformation. The current research work deals with the development of defect free Aluminium based surface nano composites, with different volume fractions of second phase particles. The work is comprised of development and deformation behaviour of aluminium based surface nano composites. Nano-structured second phases of Tungsten (W) and Aluminium oxide ($Al_2O_3$) were introduced into the Aluminium matrix using a novel unique technique, FSP. Microstructural characterization was done using SEM (Scanning Electron Microscopy) and XRD (X-ray Diffraction). Mechanical behaviour was evaluated using Vickers micro hardness. Ultimately structure – property correlations were established.

1. Introduction

Friction stir processing (FSP) is a surface modification technique being used for localized modification and control of microstructure and properties [1-3]. The basic working principle of FSP is simple and similar to Friction Stir Welding (FSW), wherein a specially designed cylindrical tool, consisting of pin and shoulder is plunged into a metal plate/block, while rotating at high speeds and is then traversed in the desired direction forming the processed zone. The processed zone results in a modified microstructure characterized by a fine and equiaxed grain structure [4].

In this current investigation, defect- free surface nano composites of Aluminium – Tungsten and Aluminium – Alumina were developed using Fiction Stir Processing (FSP). The aim behind the current experimentation was to improve the surface wear resistance of the material by developing surface composites which would be an integral part of the base material rather than acting as a detachable abrasive layer. It was assumed that the introduction of an abrasive powder into the matrix of a material would improve its surface properties and wear characteristics, in particular. Friction Stir Processing (FSP) was the technique used to homogeneously distribute the presence of these powder particles within the surface composite.

Tungsten (W), a hard metal (having Vickers hardness of 3430 MPa in solid form) with a high melting point (3422 $^o$C), taken in powder form, after being mechanically milled, was assumed to have a good resistance to wear when fabricated as Aluminium based surface nano composite using FSP. Alumina ($Al_2O_3$), an abrasive material having high hardness (Vickers hardness of 15700 MPa for 99.5 % pure material in solid form) and a high melting point (2072 $^o$C), under the similar conditions of that of W, was assumed to possess high resistance to wear [5].

Aluminium today being the most preferred material to be used in most of the engineering applications was chosen to be the composite matrix. Also being a material which could be easily worked upon, it was assumed to be a suitable matrix material to develop surface composites, using FSP [6-7]. Thus it was assumed that the integration of all these materials and techniques would produce a highly wear resistant material [8-9].

2. **Experimental Details**

The overall experimental work is divided into mechanical milling, and friction stir processing. The elemental powders of Tungsten (W) and Alumina ($Al_2O_3$), with a particle size of 10 μm and 12 μm respectively and with a mesh size of -325 mesh were used as raw materials. Both the powders were Sigma Aldrich make and had a purity of 99.9 %. The powders were handled in the glove box, which maintained an inert atmosphere of Argon (Ar). The oxygen percentage was maintained around 3.4 ppm (approx.) inside the glove box, all through the experiment. The base material used to develop surface composites was a commercially available pure aluminium block which was found to be around 94.5% pure with minor impurities of Magnesium (Mg) (0.99 %), Silicon (Si) (0.43 %) and Carbon (C) (4.08 %). The elemental composition was obtained from EDAX. The matrix material, i.e., Aluminium, was cut into blocks with dimensions 100 mm X 100 mm X 40 mm (length X breadth X height) to be used to develop surface composites using FSP.

A shaker mill (SPEX 8000), a high energy ball mill was used to mill the powders. Commercially available Friction Stir Welding machine, made by ETA Technologies, Bengaluru, was used for the experimentation purpose. A Straight cylindrical (pin length 5 mm, pin diameter 6 mm, shoulder diameter 15 mm) tool made of high carbon steel was used to develop these surface composites. A tool rotational speed of 800 rpm, a transitional speed of 40 mm/min, plunging speed of 50 mm/min and a tool tilt angle of $2^o$ were employed. The number of tool passes were varied, i.e., one tool pass and two tool passes.

During the development of surface composites using FSP, the first step was to drill holes of 2 mm diameter and 4 mm depth. Then the powders were filled into these holes and the FSP tool was passed though the Al block along the drilled holes, thus forming the surface composite layer.

The powders were characterized using SEM, XRD and TEM. The processed specimens were characterized using optical microscopy, SEM and XRD. The polished surfaces were etched using Weck's reagent.

Mechanical behaviour was evaluated using Vickers micro hardness. The surface composites were subjected to Vickers micro hardness at a load of 100 gm. The variation in hardness from the stirred zone to the base material was studied, both on the surface and on the cross section.

3. Results & Discussions

**Characterization of the Powders**

The powders of Tungsten (W) and Alumina ($Al_2O_3$) were characterized before and after milling. Figures 1 & 2 represent the XRD plots of W and $Al_2O_3$ powders and their reduction in grain size as a function of milling time is represented in the Figures 3 & 4.

From these results it can be understood that the powders of Tungsten (W) and Aluminium oxide ($Al_2O_3$) attained a reduced grain size of 5nm and 11 nm, after a milling time of 25 hrs and 32 hrs respectively. The grain size calculations were done from the XRD plots using the Scherrer formula, which is given as;

$$Grain\ size\ (t) = \frac{0.91\ \lambda}{\beta cos\theta}$$

Where,

| t | = | Grain size |
|---|---|---|
| λ | = | Wavelength of source (1.5406 A$^o$) |
| β | = | Full width half maximum |

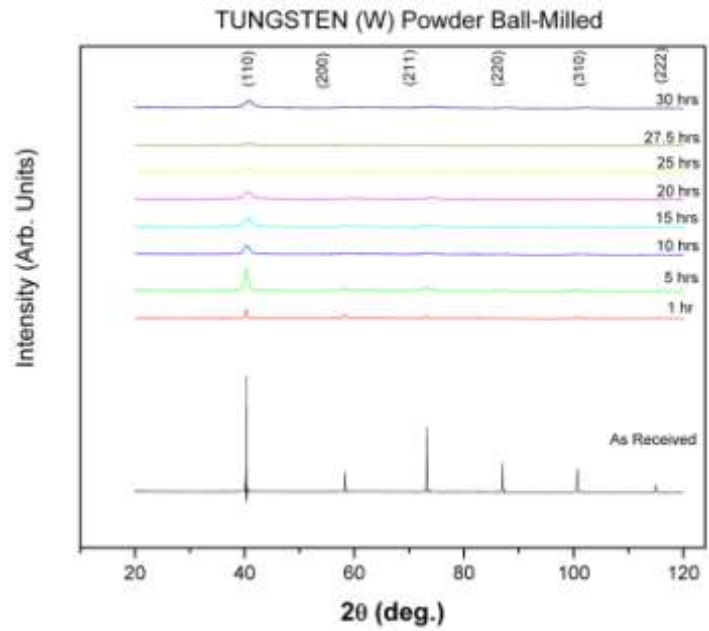

Fig 1: XRD plot of ball milled W.

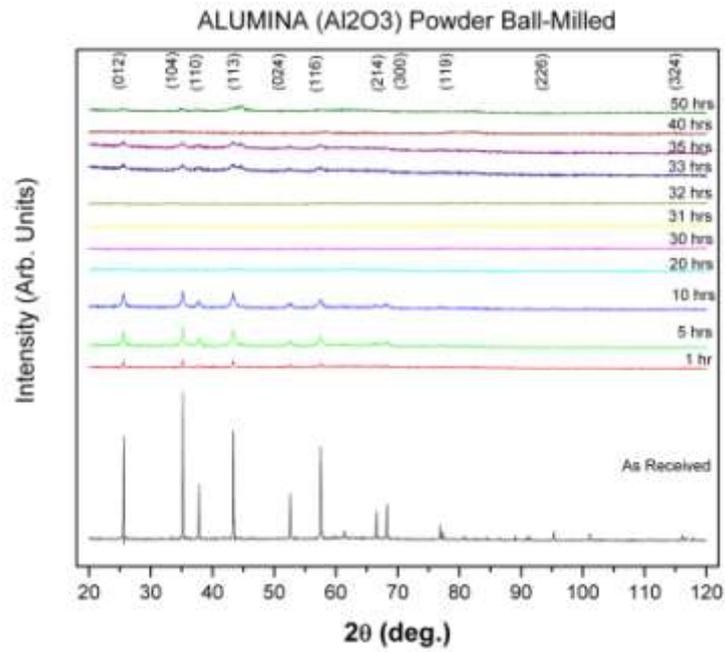

Fig 2: XRD plot of ball milled Al$_2$O$_3$.

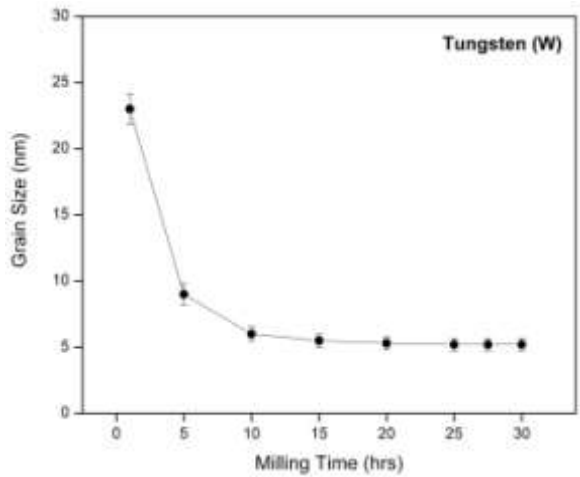 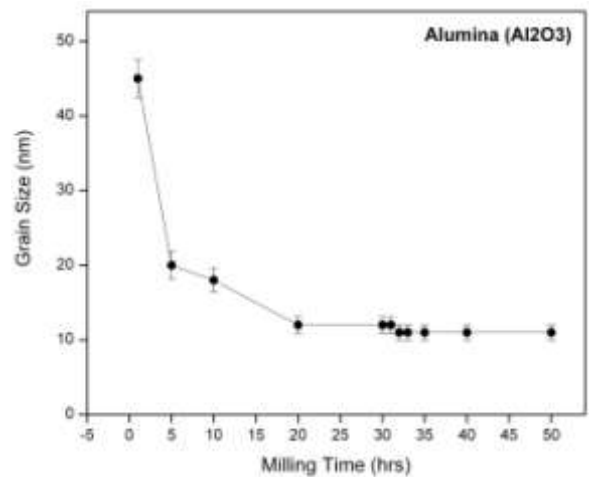

Fig 3: Milling Time vs Grain Size for W.    Fig 4: Milling Time vs Grain Size $Al_2O_3$.

The SEM results show that the powder particles agglomerated after being milled. TEM studies were conducted on the ball milled powder samples to analyze the crystallite/grain size attained by the powders after being ball milled (see Figure 5). The grain size results obtained from TEM studies approximately matched with those obtained from XRD. The EDAX results showed the presence of some alloying elements in the ball milled powders, when compared to the initially used pure powders.

**Microstructural Analysis**

The analysis of microstructure of pure aluminium shows the presence of small precipitates at the interfaces of the grain boundaries, but their presence is very minimal, which can be neglected. The average grain size measured was 7.2 μm (Fig 6). The processed material had a microstructure with a reduced grain size of 3.96 μm for single pass sample and 3.51 μm for the double pass sample. The presence of powder particles was seen on the surface of the composite. Inorder to confirm the presence of powder particles on the surface of the composite, elemental mapping was done. The elemental maps showed the presence of powder particles on the surface of the composite (Fig 7).

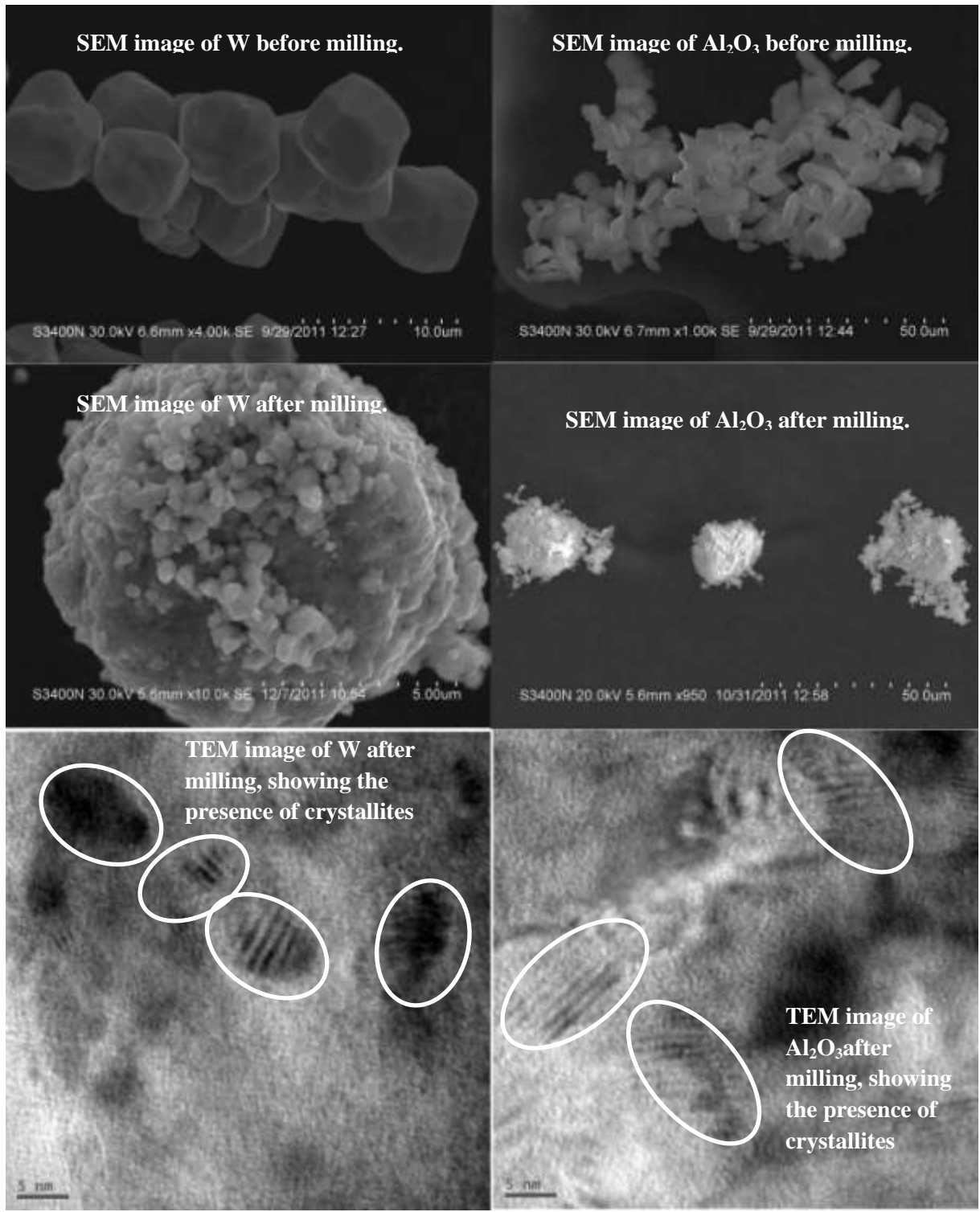

Fig 5: The SEM images of W and $Al_2O_3$ powders before and after milling. TEM images showing the presence of the nano sized crystallites formed after milling.

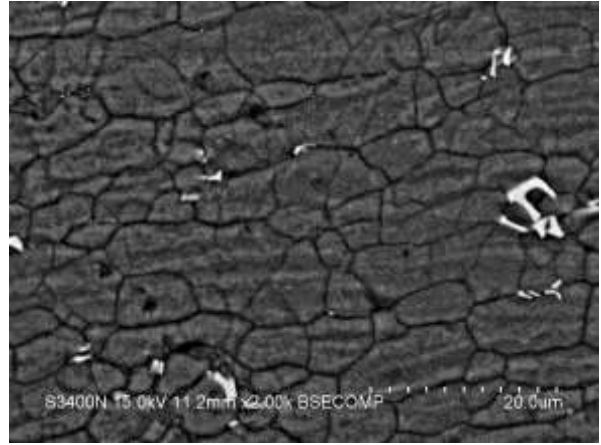

Fig 6: Microstructure of as received aluminium.

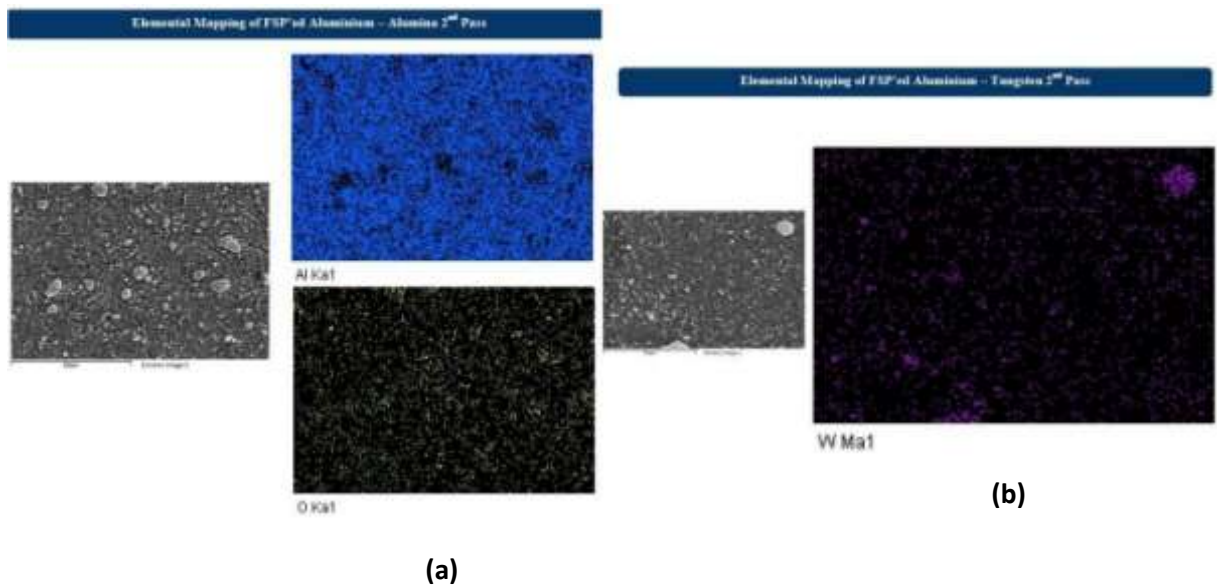

Fig 7: Elemental maps showing the presence of Alumina (a) and Tungsten (b) powders on the surface composite.

**Hardness Studies**

The hardness of pure aluminium was found to be 1047.38 MPa. Generally the hardness of the processed region increases. But in the present case, hardness of the samples reduced with an increase in the number to tool passes. The processed samples were left to cool in the open atmosphere after being processed. As aluminium is a good conductor of heat, after being processed it retained the heat and this lead to grain growth, which should rather be a grain refinement due to FSP. The grain growth occurring in place of grain refinement lead to the reduction in hardness. The stirred zone hardness comparison for all the samples, taken on the surface is shown in Figure 8.

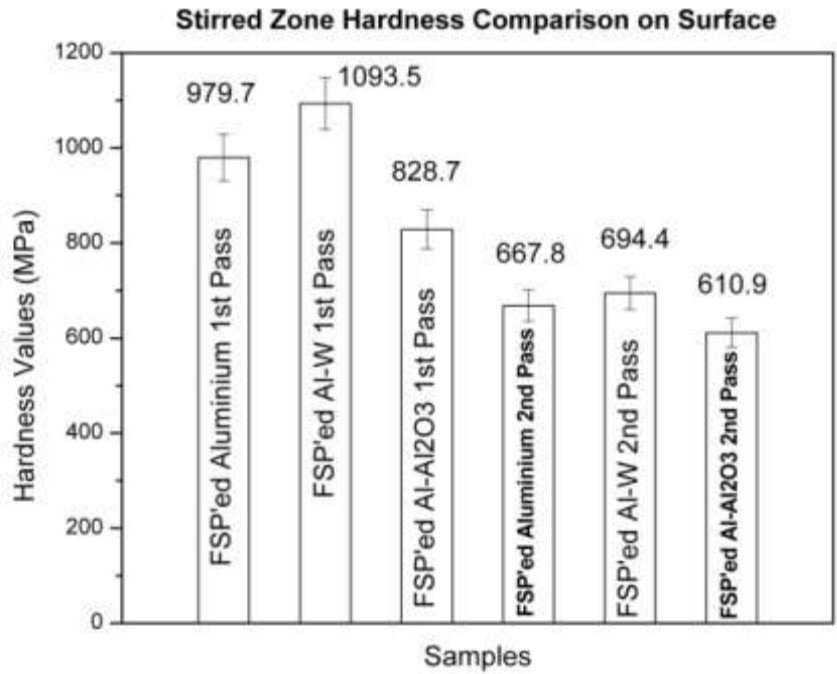

Fig 8: Stirred Zone hardness comparison on surface of all the processed samples.

4. **Summary & Conclusions**

Defect free surface nano composites of Aluminium – Tungsten and Aluminium – Alumina were developed using FSP. Powders of Tungsten (W) and Alumina ($Al_2O_3$) were milled using a Shaker Mill and grain sizes of 5 nm and 11 nm were attained after a milling time of 25 hrs and 32 hrs respectively. The second phase particles of Tungsten (W) and Alumina ($Al_2O_3$) were more 'homogeneously distributed' within the double pass samples when compared to the single pass ones. The processed specimen possessed fine grains in the processed region. The grain size of the processed specimens reduced to 3.51 μm for double pass samples and 3.96 μm for single pass samples, from an initial grain size of 7.2 μm of the base material. There was seen a reduction in the Vickers hardness values of the processed specimens. The second pass specimens had the lowest Vickers hardness values. The reason for this reduction in hardness is assumed to be a result of annealing effect, as the samples were left to cool in the open atmosphere after friction stir processing them.

# References

1. R.S. Mishra, Z.Y. Ma, "*A Review Journal on Friction Stir Welding and Processing*", Materials Science and Engineering R, 50 (2005) 19.
2. R.S. Mishra and M.W. Mahoney, "*Friction Stir Welding and Processing*", ASM International (USA), 6(2007)1.
3. Z.Y. Ma, R.S. Mishra, M.W. Mahoney, and R. Grimes, "*Friction Stir Welding of Aluminium Alloys*," Materials Science and Engineering A, 351 (2003) 148.
4. R.S. Mishra, L. Johannes, I. Charit and A. Dutta. *"Multi-Pass Friction Stir Superplasticity in Aluminium Alloys",* Proceedings of National Science Foundation, 1 (2005) 171.
5. C. Suryanarayana, "*Mechanical Alloying and Milling",* Progress in Materials Science, 46 (2001) 1.
6. E. L. Rooy, "*Introduction to Aluminium and Aluminium Alloys*", ASM Handbook, 02 (1991) 95.
7. H.J. Fecht, "*Nanostructure Formation by Mechanical Attrition*", Nanostructure Materials, 6 (1995) 42.
8. F.L. Zhang, C.Y. Wang, M. Zhu, "*Nanostructured WC/Co composite powder prepared by high energy ball milling",* Scripta Materialia, 49 (2003) 1123.
9. C.C. Koch, "*Synthesis of Nanostructured Materials by Mechanical Milling: Problems and Oppurtunities",* Nano-Structured Materials, 9 (1997) 21.